\documentclass[pra,twocolumn,amsmath,nofootinbib,superscriptaddress]{revtex4}
\usepackage{amssymb}
\usepackage{graphicx}
\usepackage{color}
\usepackage{amsfonts}

\def\be{\begin{equation}}
\def\ee{\end{equation}}
\def\bea{\begin{eqnarray}}
\def\eea{\end{eqnarray}}
\def\>{\rangle}
\def\<{\langle}

\def\qed{\leavevmode\unskip\penalty9999 \hbox{}\nobreak\hfill
     \quad$\blacksquare$
     \par\vskip3pt}
\def\bi{\begin{itemize}}
\def\ei{\end{itemize}}

\begin{document}

\title{Quantum state targeting}

\author{Terry Rudolph}
\email{rudolpht@bell-labs.com} \affiliation{Bell Labs, 600-700
Mountain Ave., Murray Hill, NJ 07974, U.S.A.}
\author{Robert W. Spekkens}
\email{rspekkens@perimeterinstitute.ca} \affiliation{Perimeter
Institute for Theoretical Physics, Waterloo,
  Ontario N2J 2W9, Canada}
\date{\today}

\begin{abstract}
We introduce a new primitive for quantum communication that we term ``state targeting'' wherein the goal is to pass a test
for a target state even though the system upon which the test is
performed is submitted prior to learning the target state's
identity. Success in state targeting can be described as having
some control over the outcome of the test. We show that
increasing one's control above a minimum amount implies an
unavoidable increase in the probability of failing the test.  This is analogous to the unavoidable disturbance to a quantum state that results from gaining information about its identity, and can be shown to be a purely quantum effect.
We provide some applications of the results to the security
analysis of cryptographic tasks implemented between remote
antagonistic parties.  Although we focus on weak coin flipping,
the results are significant for other two-party protocols, such as
strong coin flipping, partially binding and concealing bit
commitment, and bit escrow. Furthermore, the results have significance not only
for the traditional notion of security in cryptography, that of restricting a cheater's ability to bias the outcome of the
protocol, but also on a novel notion of security that arises only
in the quantum context, that of cheat-sensitivity.
Finally, our analysis of state targeting leads to some interesting secondary results, for instance, a generalization of Uhlmann's theorem and an operational interpretation of the fidelity between two mixed states.

\end{abstract}

\maketitle

\def\<{\langle}
\def\>{\rangle}
\def\be{\begin{equation}}
\def\ee{\end{equation}}
\def\bea{\begin{eqnarray}}
\def\eea{\end{eqnarray}}
\def\qed{\leavevmode\unskip\penalty9999 \hbox{}\nobreak\hfill
\quad $\blacksquare$ \par\vskip3pt}

\def\tcb#1{\textcolor{blue}{#1}}
\def\tcr#1{\textcolor{red}{#1}}

%\tableofcontents

\section{Introduction}\label{introduction}

It is well known that quantum theory allows for the implementation
of cryptographic tasks with a degree of information-theoretic
security that cannot be achieved classically, with key
distribution being the most famous example
\cite{BB84}.  Particularly interesting among these tasks are
the so-called ``post-cold-war" applications of cryptography,
wherein two remote and mistrustful parties seek to cooperate
towards some end. Examples of these are bit commitment
\cite{Mayers,LoChau,SpekkensRudolphBC}, strong coin flipping
\cite{Aharonov,Ambainis}, weak coin flipping \cite{wcf}, bit
escrow \cite{Aharonov}, and two-party secure computation
\cite{Lo}. In this paper, we focus on elucidating a certain
primitive we have identified as accounting for the ability of
certain quantum two-party cryptographic protocols to outperform
their classical counterparts.

We begin by recalling a more familiar such primitive, specifically
{\em state estimation} \cite{helstrom}: A system is prepared in
one of a set of states, and passed on to an estimator. Typically,
the task of the estimator is to try, as best as possible, to
determine which of the states describes the system. Imagine,
however, that the estimator is asked to announce his guess of the
state and then is asked to resubmit the system, which is
subsequently subjected to a pass/fail test for still being in the
initially prepared state. In this case, the estimator's task is to
guess the state without disturbing it.\footnote{Since the definition of the task of state estimation does not typically make any reference to a disturbance, we prefer to refer to
such a primitive as \emph{state spying} - it is discussed in more
detail in section \ref{statespying} }  By doing nothing to the
system, the estimator can make a random guess and avoid creating
any disturbance. However, it is well known that any interaction
with the system that leads to a greater probability of guessing
correctly will necessarily also lead to a disturbance of the state
\cite{fuchs}. This phenomena is extremely important - for instance,
it underlies the possibility of quantum key
distribution.

The focus of this paper is the task of passing a
test for a target state, even though the system upon which the
test is performed must be submitted prior to learning the target
state's identity. We call this task {\em state targeting}. State
targeting is built of the same elements as the sort of state
estimation task just discussed: There is a system and a set of
states, one of which is distinguished. There is also a classical
announcement of a state drawn from the specified set, and the most
desirable announcement is an announcement of the distinguished
state. Finally, there is a pass/fail test on the system. What is
different in state targeting is how these elements are organized,
and the fact that the person implementing the task (the player)
\emph{submits} rather than \emph{receives} the quantum system.

More explicitly, a state targeting task proceeds as follows. At
the outset, there is an unknown target state, drawn from a known
set of states. The player submits a system, and only after
doing so does she learn the identity of the target
state. At this point, the player must announce a state from the
original set (not necessarily the target state), and the system is
subjected to a pass/fail test for the announced state. Success is
defined as announcing the target state and subsequently passing
the test for this state.

The player can always make a random guess of the target state
initially, and thereafter announce this state and be sure to pass
the test for this state. This will lead to some finite probability of announcing
the target state while avoiding any risk of failing the final
test. However, it turns out that any attempt to make the
probability of announcing the target state greater than what is
achieved with this trivial scheme results in a non-zero
probability of failing the final test. This phenomena is analogous
to the unavoidable disturbance that comes with information gain in
state estimation. As one might expect, it too has interesting
applications in quantum cryptography.

The outline of the paper is as follows.  In section II, we provide
a simple cryptographic motivation for our study, namely, a weak
coin flipping(WCF) protocol wherein one of the party's cheating
strategies is an instance of state targeting. We also define the
task of state targeting in greater detail than has been done in
the introduction. In section III, we consider state targeting
where the target state is drawn uniformly from a pair of pure
states. We determine the maximum probability of success, and
investigate the rate at which the probability of failing the final
test increases with the probability of success. Section IV
generalizes the notion of state targeting by allowing for the
possibility that the player simply declines from announcing a
state.  This allows the player to achieve some non-trivial degree
of success without running any risk of failing the final test.
(This is analogous to the fact that some information gain without
disturbance becomes possible in state estimation if one has the
option to sometimes decline from resubmitting the system for
testing.) This variant of the state targeting task is also shown
to have a cryptographic motivation in terms of a weak coin
flipping protocol. In section V, we optimize this variant in the case where the target state is drawn uniformly from a pair of pure states.

In the second half of the paper, we consider state targeting for
mixed states. This requires a careful consideration of what to use
as a test for a mixed state, and so we begin in section VI by
addressing this question. In section VII, we consider the success
that can be achieved in state targeting between a pair of mixed
states, and in section VIII, we examine the case where the player
is allowed to sometimes decline from being tested.  Section IX
refines the analogy that exists between state targeting and state
estimation by focusing on a variant of state estimation, which we call state spying, wherein the estimator must resubmit the system for testing.  We consider the applications of our results on state targeting and discuss some open problems in section IX. Finally, in section X, we consider the classical
analogue of state targeting, which highlights the inherently
quantum mechanical features of this task.

\section{Motivation and definition of state targeting}\label{motivating}

To emphasize the importance of state targeting in cryptography,
and to motivate the sorts of problems that we shall address, it is
useful to have an example of a protocol that makes use of this
task.  For this purpose, we introduce a very simple protocol for
the two-party cryptographic primitive of weak coin flipping(WCF)
\cite{wcf} \cite{wcf}. Here, two separated and mistrustful
parties wish to engage in communication to generate a random bit,
the value of which will fix a winner and a loser, in such a way
that each party can be guaranteed that if they follow the honest
protocol, the other party is limited in the extent to which they
can bias the value of the bit in their favor.

The simple weak coin flipping protocol we consider makes use of a
pair of non-orthogonal pure states $|\psi_0\>$,$|\psi_1\>$.  If
both parties are honest, it proceeds as follows.

\textbf{Weak coin flipping protocol 1}
\bi
\item[1.] Alice chooses a bit $b$ uniformly from \{0,1\},
prepares $|\psi_b\>$ and sends the system to Bob.
\item[2.] Bob chooses a bit $b'$ uniformly from \{0,1\} and
announces it to Alice (one can think of this as Bob's guess of the
value of $b$).
\item[3.] Alice announces $b$ to Bob
\item[4.] Bob tests the system for being in the state
$|\psi_b\>$.

If at step (4), the system fails Bob's test, then Alice is caught
cheating. Otherwise, if $b'=b$ then Bob wins, while if $b'\ne b$,
then Alice wins.
\ei

Alice may cheat by preparing the system in an arbitrary state of
her choosing at step (1), and making whatever announcement she
pleases at step (3). Bob may cheat at step (2) by performing a
measurement upon the system, and using the outcome to inform his
decision of what $b'$ to announce. The extent to which Bob can
cheat when Alice is honest depends on the extent to which he can
correctly estimate which of a pair of non-orthogonal states
applies to the system, in order to make the best possible guess of
$b$ and maximize his probability of making $b'=b$. The problem of
state estimation has been extensively studied, and consequently
the solution can be found in the literature \cite{helstrom,fuchs}.
On the other hand, the extent to which Alice can cheat when Bob is
honest depends on the extent to which she can pass a test for the
state opposite to the one Bob guessed, despite not knowing what
Bob's guess will be when she submits the system. We can state this
as follows: her target state is determined by Bob's announcement,
which only occurs after she has submitted the system. This is a
simple instance of state targeting.

Note that Alice's announcement in step (3) of the protocol
determines what state Bob is to test for.  Alice can only win the
coin flip if she announces a bit $b$ that is unequal to $b'$.
However, if she announces $b\ne b'$ without having initially
prepared $|\psi_b\>$, then she runs a risk of failing Bob's test.
Of course, Alice may sometimes prefer to pass a test for the
non-target state rather than failing the test for the target
state.  As such, she may not always ask Bob to test for the target
state.

The task faced by a dishonest Alice in our simple WCF protocol is
just one instance of state targeting. More generally, a state
targeting task is defined as follows:
\bi
\item[(i)] Alice submits a system to Bob.
\item[(ii)] Alice learns the identity of the target state.
\item[(iii)] Alice announces a state to Bob (not necessarily
the target state).
\item[(iv)] Bob performs a Pass/Fail test for the announced
state.
\ei

The possible outcomes are:
\begin{itemize}
\item[(A)] Alice announces the target state and passes Bob's test
\item[(B)] Alice announces the target state and fails Bob's test
\item[(C)] Alice announces a non-target state and passes Bob's test
\item[(D)] Alice announces a non-target state and fails Bob's test
\end{itemize}

``Success" in state targeting is to achieve outcome (A). Thus, we
can quantify the degree of success by the probability of this outcome. We call this
probability Alice's {\em control} \footnote{Note that in the case of two target states with equal prior probability, our definition of this quantity coincides with the definition used in Ref.~\cite{SpekkensRudolphBC} except that it is offset from the latter by a factor of 1/2.}  We shall be interested in
determining the maximum control achievable in state targeting.

There are several ways in which Alice might not succeed.  In
cryptographic applications, it is reasonable to expect that
failing one of Bob's tests has a greater cost than passing the
test for a non-target state, since the former indicates
that Alice has cheated. Thus, it is useful to consider the total probability
of failing Bob's test, which is the sum of the probabilities for
outcomes (B) and (D).  We call this probability Alice's {\em
disturbance}. It is sometimes useful for Alice to sacrifice some
control to lower her disturbance. Thus, we shall be interested in
determining the minimal disturbance for a given control, which we
refer to as the optimal \textit{control-disturbance trade-off}.

\section{State targeting for two pure states} \label{ST2PS}

\subsection{Maximum control} \label{ST2PSmaxcontrol}

Consider the example of state targeting that is provided by our
simple weak coin flipping protocol, namely, one wherein the target
is selected uniformly from a pair of pure states,
$|\psi_0\>,|\psi_1\>$.  If Alice simply wishes to maximize her
probability of announcing the target state and passing Bob's test,
that is, if she is unconcerned about the relative probabilities of
outcomes (B),(C) and (D), then she should always announce the
target state.  Her most general strategy at step (i) is to prepare
the system in a (possibly mixed) state $\rho$. In this case,
Alice's control is
\be
C=\frac{1}{2}\<\psi_0|\rho|\psi_0\>+\frac{1}{2}\<\psi_1|\rho|\psi_1\>.\ee
Noting that this can be rewritten as $\frac{1}{2}{\rm Tr}(\rho
(|\psi_0\>\<\psi_0|+|\psi_1\>\<\psi_1|)$, it is clear that the
maximum control is simply the maximum eigenvalue of
$|\psi_0\>\<\psi_0|+|\psi_1\>\<\psi_1|$ and is achieved by
choosing $\rho$ to be the associated eigenvector.  Explicitly,
the maximum control is
\begin{equation}  \label{maxcontrol}
C^{\max }\equiv\max_{\rho}\, C= \frac{1}{2}(1+|\langle \psi
_{0}|\psi _{1}\rangle |)
\end{equation}
and the state that achieves it is
\be |\psi_M\>\equiv {\cal N}(|\psi_0\>+|\psi_1\>), \label{psiopt4control} \ee
where ${\cal N}$ is a normalization factor, and the phases have
been chosen so that $\<\psi_0|\psi_1\>$ is real.

\begin{figure}
\includegraphics[width=60mm]{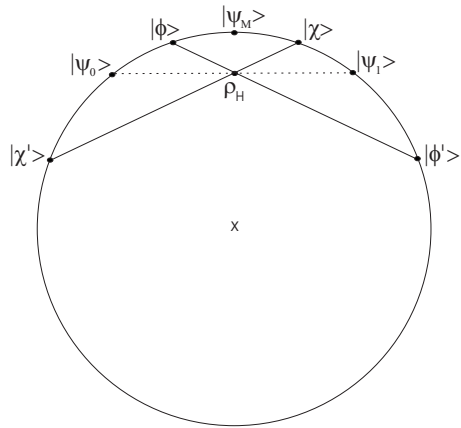}
\caption{A slice of the Bloch sphere, showing the two possible target states,
$|\psi_0\>$ and $|\psi_1\>$. The state $|\psi_M\>$ is the one that Alice must submit to
achieve her maximal control. If she submits the state
$\rho_H$, then she can achieve some non-trivial control at the cost of some disturbance by steering to a particular convex decomposition of the state prior to her announcement. The pair of states
$|\phi\>,|\phi'\>$ indicate the elements of a convex decomposition that Alice might use where the target state is  $|\psi_0\>$. In general the optimal state
to submit in order to minimize her disturbance for a given control lies between $\rho_H$ and
$|\psi_M\>$.}\label{Blochmaxcontrol}
\end{figure}

 Since the Hilbert
space spanned by any pair of pure states is two-dimensional, it
can be represented using the Bloch sphere picture. Within this
picture, $|\psi_M\>$ is represented by the point that lies halfway
between the points representing $|\psi_0\>$ and $|\psi_1\>$ on the
geodesic that connects them, as indicated in
Fig.~\ref{Blochmaxcontrol}.

\subsection{The control-disturbance trade-off}\label{CversusD}

By achieving her maximum control, Alice runs a risk of failing
Bob's test.  Since she always announces the target state in this
case, outcomes (C) and (D) never occur and her probability of failing Bob's test (her disturbance) is
simply the difference between 1 and her maximum control, that is,
$\frac{1}{2}(1-|\langle \psi _{0}|\psi _{1}\rangle |)$. If, on the
other hand, Alice exerts her trivial control of 1/2 by preparing
$|\psi_0\>$ or $|\psi_1\>$ with equal probability and then
announcing this state regardless of the identity of the target
state, she creates no disturbance. We can imagine, however, that
situations may arise wherein Alice decides to achieve some non-trivial but non-maximal control. It is useful therefore to
determine the minimal disturbance for every achievable degree of
control. Since we find that this function is monotonically
increasing, it also specifies the maximum control for a given
disturbance.

 We begin by making the observation that a second strategy for Alice which
achieves the trivial control with no disturbance is for her to
couple the system she sends to Bob with an ancilla that she keeps,
thereby preparing the entangled state
\begin{equation}  \label{honestrho}
\frac{1}{\sqrt{2}}\left(|0\>|\psi_0\>+|1\>|\psi_1\>\right).
\end{equation}
When it comes time to announce a state to Bob, she simply measures
the ancilla in the $|0\>, |1\>$ basis, registers the bit value $b$
of the outcome, and announces $|\psi_b\>$.

In this situation the reduced density operator for the submitted
system is $\rho_H = \frac{1}{2}| \psi_0 \rangle \langle \psi_0
|+\frac{1}{2}| \psi_1 \rangle \langle \psi_1 |$, which is
represented on the Bloch sphere by the point which lies halfway
along the line joining the points corresponding to $|\psi_0\>$ and
$|\psi_1\>$ - this is depicted in Fig.~\ref{Blochmaxcontrol}.

Alice can now use the following trick to increase her control
beyond the minimum value. Suppose Bob announces that the target
state is $|\psi_0\>$. Alice then implements a measurement on her
ancilla of some basis distinct from the ${|0\>,|1\>}$ basis. This
collapses the state of the submitted system to $|\phi\>$ with
some probability $q$ and to $|\phi'\>$ with probability $(1-q$).
An example is depicted in Fig.~1. We assume that of the two
states, $|\phi\>$ has the greater overlap with $|\psi_0\>$ (as is
the case in the figure), so that Alice announces $|\psi_0\>$ upon
obtaining the outcome associated with $|\phi\>$, and she announces
$|\psi_1\>$ upon obtaining the outcome associated with $|\phi'\>$.
If the target state is $|\psi_1\>$, she exerts a similar strategy,
collapsing the state of the submitted system to a different pair of states, $|\chi\>$
or $|\chi'\>$, with probabilities $p$ and $1-p$
respectively.\footnote{The possibility of influencing which of
several convex decompositions describe the updating of the state
of a remote system by choosing which measurement to perform on a
system with which it is entangled is the key element of the EPR
argument
\cite{EPR}. Schr\"{o}dinger described this phenomena as
\textit{steering} the state of a remote system \cite{schr32}. We adopt this term, but are careful to avoid saying that the state is steered, since what one can steer between are different convex decompositions of the state.}

By this strategy, Alice achieves a control of \be\label{ccc}
C=\frac{1}{2} q|\<\phi|\psi_0\>|^2 + \frac{1}{2}
p|\<\chi|\psi_1\>|^2. \ee Her disturbance, which is her total
probability of failing Bob's test, is
\bea
D&=&\frac{1}{2}[q(1-|\<\phi|\psi_0\>|^2)+(1-q)(1-|\<\phi'|\psi_1\>|^2]
\nonumber\\
&+&\frac{1}{2}[p(1-|\<\chi|\psi_1\>|^2)+(1-p)(1-|\<\chi'|\psi_0\>|^2]
\eea

One might worry that of all the convex decompositions of $\rho_H$,
only some can be realized by performing a measurement on the
ancilla.  If this were the case, then this subset of convex
decompositions would have to be characterized before we could
proceed with the optimization. However, this is not a concern,
because as is shown in Refs. \cite{schr32} and \cite{HJW} (and generalized to
non-extremal decompositions in \cite{RudolphSpekkensQIC}),
\textit{every} convex decompositions of a mixed state can be
achieved by some measurement on the ancilla.  Thus, we can vary
over all convex decompositions with the assurance that there is a
measurement that will achieve it.
 Assuming $|\psi_0\>$ and $|\psi_1\>$ are non-orthogonal, $\rho_H$ lies off
the centre of the Bloch sphere, and consequently $|\phi\>$ and
$|\phi'\>$ (which define the convex decomposition of $\rho_H$) can
be chosen such that $q>1/2$; this follows from the fact that $\rho_H$ is geometrically closer to $|\phi\>$
than to $|\phi'\>$ in the Bloch sphere. Similarly, Alice can ensure that $p>1/2$.
Despite the fact that Alice has a non-unit probability of passing
Bob's test, her overall probability of success can increase.

\begin{figure}
\includegraphics[width=90mm,clip=]{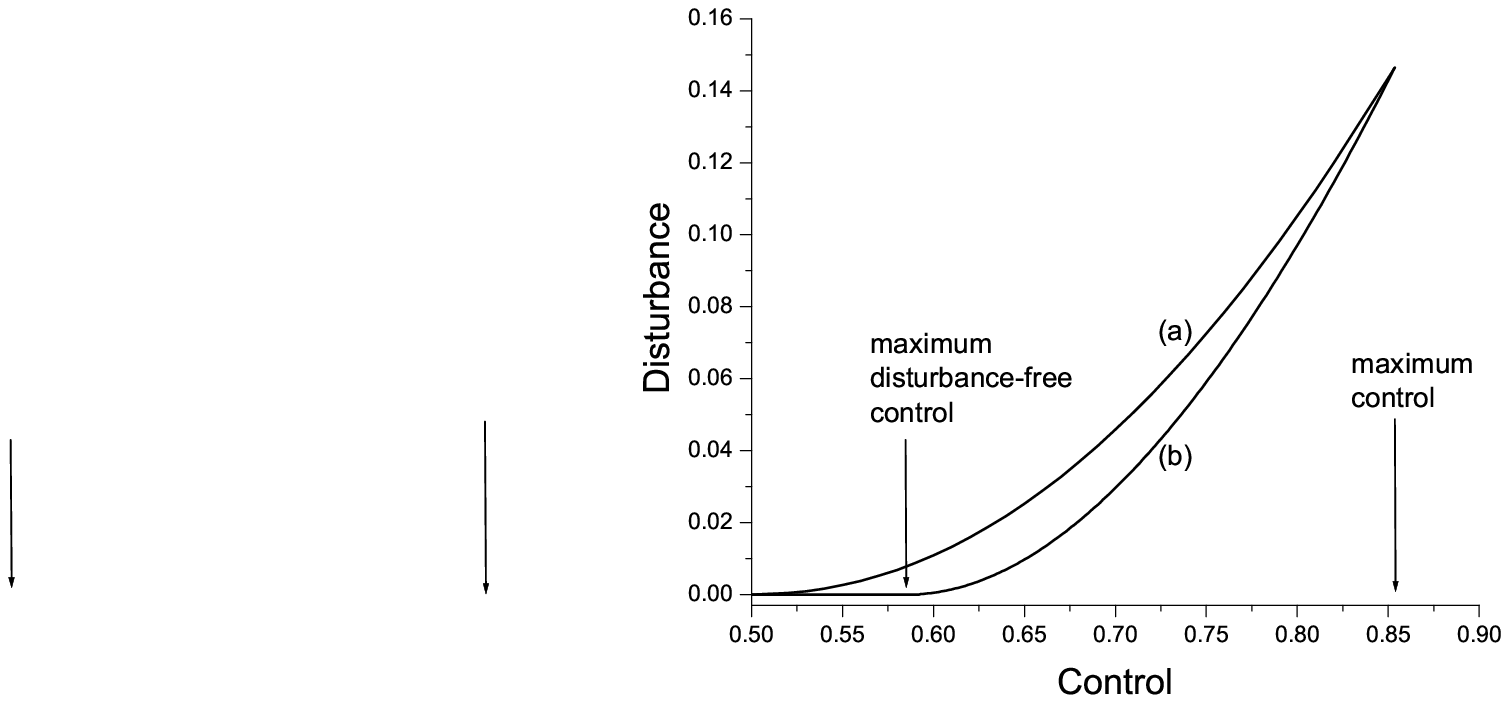}
\caption{The control disturbance tradeoff for two pure states, with $|\<\psi_0|\psi_1\>|^2=1/2$.
Curve (a) is for the case where Alice does not have the option to decline from being tested, considered in section
\ref{CversusD}, whereas curve (b) is for the case where she does, considered in section \ref{dfCversusD}.  For the former, the disturbance is non-zero for every control above 1/2, while for the latter it is only non-zero after the point indicated in
the diagram. }\label{tradeoff}
\end{figure}

We do not expect that $\rho_H$ will correspond to the optimal
$\rho$ for every value of the control that Alice may wish to
achieve. Indeed, we already know that in order to achieve her
maximum control, she must submit a different state, indicated by
$|\psi_M\>$ in Fig.~\ref{Blochmaxcontrol}. Thus we expect the
optimal $\rho$ to be an interpolation between $\rho_H$ and
$|\psi_M\>$ as
we increase $C$ from $1/2$ to $C^{\max}$. So, we must perform
two optimizations: an optimization over the state $\rho$ of the
submitted system, and an optimization over the convex
decomposition of $\rho$ she should realize prior to announcing to
Bob which state he should test for.

It is not difficult to show that the optimal $\rho$ has a Bloch
vector that bisects the Bloch vectors representing $|\psi_0\>$ and
$|\psi_1\>$, and that the optimal convex decomposition to realize
when $|\psi_1\>$ is the target state is simply the mirror image in
the Bloch sphere of the optimal convex decomposition when
$|\psi_0\>$ is the target state, so that $q=p$. These
simplifications leave us with a two parameter optimization
problem: the one parameter corresponding to the length $r$ of the
Bloch vector representing $\rho$, the other parameter
corresponding to the angle between the Bloch vectors representing
$|\psi_0\>$ and $|\phi\>$. Although, we have not been able to
perform this optimization analytically for very value of $|\<\psi_0|\psi_1\>|$. The curve~(a) in Fig.~\ref{tradeoff} depicts the optimal tradeoff
for $|\<\psi_0|\psi_1\>|^2=1/2$.

\section{Motivation and definition of disturbance-free state
targeting} \label{dfstatetargeting}

In the context of the WCF protocol presented earlier, it may sometimes be the case that the consequences
for Alice if she is caught cheating are sufficiently dire that she
wishes to avoid this outcome at all costs. The question then
becomes whether this is enough to guarantee that she follow the
honest protocol. For the
case of two non-orthogonal target states, if Alice submits some state distinct from one of
these possibilities, then regardless of what she announces, she
has a non-zero probability of failing Bob's test.  The only way to
be certain to pass Bob's test is to submit one of these states and
to announce the same state, that is, to follow the honest protocol and achieve the trivial control of 1/2. This is reflected in curve (a) of Fig.~(\ref{tradeoff}) by the fact that for every value of control greater than 1/2 the disturbance is non-zero.

Nonetheless, there are useful versions of the task of state
targeting wherein Alice has the option of declining from
announcing a state to Bob, and in such cases, Alice {\em can}
achieve greater than the trivial control without incurring any
disturbance.

We again introduce a simple WCF protocol as motivation. This
differs from the previous one in that both Alice and Bob are
tested for cheating, which results in a protocol that offers fewer
cheating possibilities for Bob, but more for Alice. If both
parties are honest, the protocol proceeds as follows.\\

\textbf{Weak coin flipping protocol 2}
\begin{itemize}
\item[1.] Alice chooses a bit $b$ uniformly from \{0,1\},
prepares $|\psi_b\>$ and sends the system to Bob.
\item[2.] Bob chooses a bit $b'$ uniformly from {0,1} and
announces it to Alice (one can think of this as Bob's guess of
which state Alice submitted).
\item[3.] Alice announces $b$ to Bob

\textbf{If} $b'=b$, then
\item[4.] Bob returns the system to Alice, and Alice tests the
system for being in the state $|\psi_b\>$.
\noindent

\textbf{Else}, if $b'\ne b$, then
\item[4.] Bob tests the system for being in the state
$|\psi_b\>$.

If at step (4) the system fails Alice's(Bob's) test, then
Bob(Alice) is caught cheating. Otherwise, if $b'=b$ then Bob wins,
while if $b'\ne b$, then Alice wins.
\end{itemize}

As before, Alice may cheat by preparing the system in an arbitrary
state of her choosing, and making whatever announcement she
pleases. The difference is that if she announces $b'=b$, then she
is not tested by Bob.

This suggests a more general type of state targeting task, which
differs from the one outlined in section \ref{motivating} insofar
as step (iii) is replaced by:
\bi
\item[(iii${}^\prime$)] Alice has the option of either (a) announcing a
state to Bob (not necessarily the target state), or (b) declining
to announce a state to Bob.
\ei
There is also an additional possible outcome relative to the
version from the last section, namely:
\bi
\item[(E)] Alice declines to announce a state to Bob.
\ei

Because of this additional outcome, Alice can make her disturbance
strictly zero while still achieving a non-trivial control.  We
call this {\em disturbance-free state targeting}.  The control
that Alice achieves by implementing her best disturbance-free
state targeting is still defined as her probability of achieving
outcome (A).
We call this her {\em disturbance-free control}, and denote it by
$C_\textrm{df}$. (An analogous quantity is defined in Ref.
\cite{wcf}, where it is called Alice's {\em threshold for
cheat-sensitivity}.)

The possibility of disturbance-free state targeting is analogous to something that occurs in \emph{unambiguous discrimination} of pure states\cite{IDP}. (The latter is a procedure which either discriminates the states without any probability of error, or else simply returns the result ``inconclusive''.)  If the inconclusive outcome is \emph{not} obtained, then the pure state with which the system was prepared is known with certainty, and can be re-prepared, so that a test for the initial state can be passed with certainty.  If the inconclusive result \emph{is} obtained, then the estimator declines from having the system tested.   Thus, the existence of an inconclusive outcome allows for the possibility of information gain without disturbance.  Analogously, the option to decline from being tested allows for non-trivial control without disturbance.

\section{Disturbance-free state targeting for two pure states}

\subsection{Maximum disturbance-free control}
We now demonstrate how disturbance-free state targeting is
achieved in the case where the target state is chosen uniformly
from two non-orthogonal pure states. The technique is similar to
the one used to minimize the disturbance for a given control.
Alice initially couples the submitted system to an ancilla (which
she keeps) so that the state of the pair is entangled. After
learning the identity of the target state, she measures her
ancilla in such a way that the state of the submitted system
is updated according to a convex decomposition that contains
the target state. If it happens to collapse to the target state,
then she announces the target state, and is certain to pass Bob's
test.  If it collapses to another state, then she simply declines
to announce a state to Bob.

\begin{figure}
\includegraphics[width=70mm]{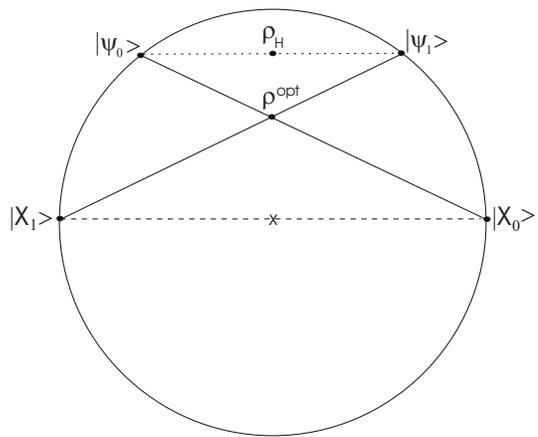}
\caption{A section of the Bloch sphere showing the state $\rho ^{\mathrm{opt}}$ that yields
the largest disturbance-free control when the possible target states are $|\psi_0\>$ and $|\psi_1\>$.}\label{UnambigBloch}
\end{figure}

%In the Bloch sphere, disturbance-free state targeting %corresponds to submitting a state $\rho$ and realizing convex %decompositions of $\rho$, one element of which
The fact that such a scheme can achieve a control greater than 1/2 can be seen by considering Fig.~(\ref{UnambigBloch}).  If the state $\rho$ that Alice submits falls between $\rho_H$ and the completely mixed state (the centre of the Bloch
sphere), and she steers to a two-element convex decomposition containing the target state, then because $\rho$ is necessarily geometrically closer to the target state than to
the other element of the decomposition, and because greater geometric proximity represents a greater probability of collapsing to that element \cite{RudolphSpekkensQIC}, it follows that the probability of announcing the target state is necessarily greater than 1/2.

We now proceed to find the maximum probability of disturbance-free
state targeting.  We must vary over both the convex decompositions
for a fixed submitted state, and over the submitted states.

In the first optimization, we seek to
find the largest probability with which the target state
appears in a convex decomposition of the submitted state. A
corollary we prove in section (\ref{dfst2mixed}) establishes that
the maximum probability of collapsing the state $\rho$ to the
state $|\psi\>$ is
\be
\frac{1}{\left\langle \psi \right| \rho ^{-1}\left| \psi
\right\rangle }.
\ee
Given that the target state is equally likely to be $|\psi_0\>$ or
$|\psi_1\>$, the disturbance-free control for a submitted state
$\rho$ is \be \label{expression4dfcontrol} C_{\text{df}}
=\frac{1}{2} \frac{1}{\left\langle \psi _{0}\right| \rho
^{-1}\left| \psi _{0}\right\rangle }+\frac{1}{2}\frac{1}{
\left\langle \psi _{1}\right| \rho ^{-1}\left| \psi
_{1}\right\rangle } \ee

We can now consider the optimization over $\rho$. Since there is
clearly no advantage to preparing a $\rho $ outside of the
subspace spanned by $\left| \psi _{0}\right\rangle $ and $\left|
\psi _{1}\right\rangle ,$ the problem can be entirely formulated
in a two-dimensional Hilbert space, so that we may use the Bloch
sphere representation of states. The optimization is done in
Appendix \ref{dfcontrol}. The maximum disturbance-free control is
found to be
\be\label{maxdfcontrol} C^{\text{max}}_{\text{df}}
\equiv \max_{\rho}C_{\text{df}} = \frac{1}{1+\sqrt{1-\left|
\left\langle \psi _{0}|\psi _{1}\right\rangle \right| ^{2}}}.\ee

The strategy that achieves this control requires Alice to prepare
a mixed state of the form
\begin{equation}
\rho ^{\mathrm{opt}}=\frac{\alpha}{2}I+\frac{1-\alpha}{2}\left(
\left| \psi _{0}\right\rangle \left\langle \psi _{0}\right|
+\left| \psi _{1}\right\rangle \left\langle \psi _{1}\right|
\right) , \label{opt rho for unamb C}
\end{equation}
where
\be
\alpha=1-\frac{1}{2}\frac{1}{\left| \left\langle \psi _{0}|\psi
_{1}\right\rangle \right| ^{3}}\left( 1-\sqrt{1-\left|
\left\langle \psi _{0}|\psi _{1}\right\rangle \right| ^{2}}\right)
.
\ee

It is easy to see that the maximum disturbance-free control,
Eq.~(\ref{maxdfcontrol}), is less than the maximum control,
Eq.~(\ref{maxcontrol}), for any pair of states. Thus, to avoid
creating a disturbance, Alice must pay a price in control.

%\tcb{
%Also, perhaps a good point to mention the analgous thing in %state
%estimation???}

%In terms of the Bloch sphere picture, $\rho ^{\mathrm{opt}}$ has a
%particularly simple relation to $\left| \psi _{0}\right\rangle $
%and $\left| \psi _{1}\right\rangle $. Suppose $\hat{a}_{0}$ and
%$\hat{a}_{1}$ are the Bloch vectors associated with $\left| \psi
%_{0}\right\rangle $ and $\left| \psi _{1}\right\rangle ,$
%and that $\hat{d}=\frac{\hat{a}_{0}-\hat{a}_{1}}{\left| \hat{a}_{0}-\hat{a}%
%_{1}\right| }.$ The point on the Bloch sphere associated with
%$\rho^{\mathrm{opt}}$ is the intersection between the line segment
%with endpoints $\hat{a}_{0},-\hat{d}$ and the line segment with
%endpoints $\hat{a}_{1},\hat{d }.$ This is depicted in
%Fig.~\ref{UnambigBloch} by the point $U$. The points $\hat{d}$ and
%$-\hat{d}$ represent the states that Bob's system is collapsed to
%when Alice declines to announce a state.  For instance, if the
%target state is $|\psi_0\>$, then Alice realizes the convex
%decomposition of $\rho^{\mathrm{opt}}$ associated with the line
%segment with endpoints $\hat{a}_{0},-\hat{d}$.

In terms of the Bloch sphere picture, $\rho ^{\mathrm{opt}}$ has a
particularly simple geometric relation to $\left| \psi _{0}\right\rangle $
and $\left| \psi _{1}\right\rangle $, as is depicted in
Fig.~\ref{UnambigBloch}. Note that the states $|X_0\>$ and
$|X_1\>$ are orthogonal and symmetric about $|\psi_0\>$ and $|\psi_1\>$. If the target state is $|\psi_0\>$, then Alice steers to the decomposition of $\rho^{\mathrm{opt}}$ containing $|\psi_0\>$ and $|X_0\>$.

\subsection{The control-disturbance trade-off for state targeting
with an option to decline}\label{dfCversusD}

 It is interesting to determine
the minimal disturbance for a given control when there is an
option to decline from announcing a state. In this case, the
disturbance will be non-zero only for controls greater than the
disturbance-free control. However, the disturbance for the maximal
control is the same, since to achieve her maximal control, Alice
must always announce the target state, and consequently never
exercises her option to decline from announcing a state.

We consider this tradeoff in the simple case  of two pure states.
Suppose that Alice has submitted some $\rho$ and now
learns that the target state is $|\psi_0\>$. As described in section \ref{CversusD}, she can
make a measurement on the ancilla which collapses the submitted
system to either the state $|\phi\>$ (with probability $q$) or to
state $|\phi'\>$ (with probability $1-q$). As before, if she
collapses to the state $|\phi\>$ then she announces the state
$|\psi_0\>$. Unlike before however, if she collapses $\rho$ to
$|\phi'\>$ then she declines to announce a state to Bob rather
than announcing $|\psi_1\>$. A similar strategy is adopted if the
target state is $|\psi_1\>$. This scenario does not affect the
expression for the control that Alice achieves, it is still given
by (\ref{ccc}). However her disturbance is now given by \be
D=\frac{1}{2}q(1-|\<\phi|\psi_0\>|^2)+\frac{1}{2}p(1-|\<\chi|\psi_1\>|^2)
\ee

 Once again the problem of minimizing Alice's disturbance for a given control can
be simplified to an optimization over only two parameters. The
solutions to this optimization problem can be found analytically.
Specifically, we obtain the optimal control and disturbance as
parametric equations in a parameter $\mu$ which runs from 0 to
$\theta$, where $\theta=|\<\psi_0|\psi_1\>|$ by definition. Explicitly:
\begin{eqnarray}
C^{opt}&=&\frac{1}{2}
\frac{\cos\mu-\sin(\theta-\mu)}{1+\sin(\theta-\mu)} \\
D^{opt}&=&\frac{1}{2} \frac{1-\cos\mu}{1+\sin(\theta-\mu)}
\end{eqnarray}
The length $r$ of the Bloch vector corresponding to the optimal
$\rho$ that Alice should submit for a given control is \be
r^{opt}=\frac{1-\sin(\theta-\mu)}{\cos(\theta-\mu)} \ee Curve (b) of Fig.~(\ref{tradeoff}) is a plot
of this optimal tradeoff for $\theta=\pi/4$.

\section{Testing for a mixed state}\label{tests}

We have thus far examined state targeting under the assumption
that the possible target states are pure. A useful and powerful
generalization of state targeting occurs if we admit the
possibility of mixed target states. Recall that the task of state
targeting requires Bob to test the submitted system for being in
the state announced by Alice. Thus, we must begin by discussing
the possible ways in which one can test for a mixed state. Because
greater difficulty in passing a test typically yields greater
security for the
 protocol making use of it, we seek to find the tests
that are most difficult to pass.

For convenience, we call the tester Bob and the one being tested
Alice. The only kind of measurement
on the submitted system alone that is always passed when the system is in the state
$\sigma$ is one that is associated with a projector onto a
subspace that is greater than or equal to the support of $\sigma.$
The most difficult such measurement to pass is the one associated with the
projector onto the support of $\sigma$. We call this the
\textit{support test} for $\sigma$. Unfortunately, there are many
other mixed states which always yield a positive outcome for this
test, namely those having the same support as $\sigma.$ Thus, this
test is relatively easy to pass.

Another interesting test is one that makes use of a particular
convex decomposition of $\sigma$, that is, a set
$\{p_k,|\psi_k\>\}$ such that $\sum_k
p_k|\psi_k\>\<\psi_k|=\sigma$.  The idea is that if Alice prepared
the system in a state by drawing it from the set $\{|\psi_k\>\}$
according to the distribution ${p_k}$, then from Bob's perspective, she has submitted the state $\sigma$. If
Alice does not implement such a preparation, then it is possible for Bob to
sometimes detect this fact; he simply demands that Alice announce
the value of $k$ to him, and he then tests the system for being in
the state $|\psi_k\>$. We call this a {\em convex decomposition
test} for $\sigma$.  This test is better than the support test,
because if Alice initially prepares a mixed state that merely has
the same support as $\sigma$, this does not guarantee that she can
pass a convex decomposition test with certainty.  Nonetheless, it
is still a weak test since there any mixed states besides $\sigma$
that can pass the test with certainly, namely, any state that is a convex sum of
the ${|\psi_k\>}$.  The problem is that the test cannot verify
anything about the probability distribution from which the state
was drawn.  Convex decomposition tests arise often in the design
of cryptographic protocols
\cite{BB84,Aharonov,Ambainis,RudolphSpekkensQIC}, although they
are not typically identified as tests {\em for} a particular mixed
state.

 The best way of testing whether a system is described by a
state $\sigma$ is to test whether a larger system, of which the
first is a subsystem, is in a purification of $\sigma$. Suppose
Alice initially submits a system $B$ to Bob. When it comes time to
test this system for being in $\sigma$, Alice is required to
submit a second system, called the message system $M$, and Bob
measures the pair $MB$ for being in a particular purification of
$\sigma$, that is, for being in a state
$|\Psi\>$ such that ${\rm
Tr}_M(|\Psi\>\<\Psi|)=\sigma$ (there are
many purifications of a mixed state, but all these yield
equivalent tests). This is better than either the support or
convex decomposition tests. First of all, unlike the latter tests,
Alice can only pass this test with certainty if the state of the
submitted system was $\sigma$. More importantly, it can be shown
that for every convex decomposition test, there is a purification
test that is equally or more difficult to pass. This is proven in
appendix \ref{convspur}.

Notwithstanding this fact, in the context of state targeting
between a pair of mixed states, it is often possible to find a
pair of convex decomposition tests that yield a maximum control as
low as the maximum control for a pair of purification tests. An
example is offered by the coin flipping protocols proposed in
Ref.~\cite{Ambainis} and in Ref.~\cite{SpekkensRudolphBC}, where
Alice's optimal cheating strategies are state targeting procedures that
achieve the same control. Because of the practical
difficulty of generating, preserving and measuring entangled
states, it is likely that, other things being equal, a pair of
convex decomposition tests is preferable to a pair of purification
tests for use in a cryptographic protocol. However, it is an open
question whether for \emph{every} pair of purification tests there
is a pair of convex decomposition tests that yield the same
maximum control. In any event, it is clear that convex
decomposition tests cannot do better than purification tests and
therefore for the rest of this paper we focus on the latter.

\section{State targeting for two mixed states}\label{maxmixedcontrol}

%Take $B$,$M$,$E$ to token $t$, proof $p$ and ancilla $a$.

We begin by motivating the generalization of state targeting to
mixed states by describing another improvement on our simple weak
coin flipping protocol.  The protocol is defined in terms of a
pair of states $|\Psi_0\>$ and $|\Psi_1\>$ on a bi-partite system,
where the parts are denoted by $M$ and $B$. If both parties are
honest, it proceeds as follows.

\textbf{Weak coin flipping protocol 3}
\bi
\item[1.] Alice chooses a bit $b$ uniformly from \{0,1\}, prepares the
system $MB$ in the state $|\Psi_b\>$ and sends $B$ to Bob.
\item[2.] Bob chooses a bit $b'$ uniformly from \{0,1\} and
announces it to Alice (one can think of this as Bob's guess of
$b$).
\item[3.] Alice announces $b$ and sends the system $M$ to Bob.
\item[4.] Bob tests the system $MB$ for being in the state
$|\Psi_b\>$.

If at step (4) the system $MB$ fails Bob's test, then Alice is
caught cheating. Otherwise, if $b'=b$ then Bob wins, while if
$b'\ne b$, then Alice wins.
\ei

 Assuming the states $|\Psi_0\>$ and $|\Psi_1\>$ are
entangled over ${\cal H}^M \otimes {\cal H}^B$, their reduced
density operators on $B$, denoted $\sigma_0$ and $\sigma_1$, are
mixed. $|\Psi_b\>$ is a purification of
$\sigma_b$.  Thus, Bob's test at step (4) is a purification test
for $\sigma_b$. It should now be clear that Alice's task when
cheating is an instance of state targeting with two mixed target states, since the state $\sigma_b$ that
Alice would like to pass a test for is determined by Bob's
announcement, which occurs after she has submitted system $B$.

Note that if $|\Psi_0\>$ and $|\Psi_1\>$ are taken to be product states, this protocol reduces to WCF protocol 1. Making use of entangled $|\Psi_0\>$ and $|\Psi_1\>$ (equivalently, mixed $\sigma_0$ and $\sigma_1$) provides an improvement over WCF protocol 1 insofar as it provides better simultaneous security against Alice and Bob. This is shown explicitly in Ref.~\cite{SpekkensRudolphBC}. Briefly, it follows from the fact that the security against Alice is quantified by the fidelity between $\sigma_0$ and $\sigma_1$ (as we shall show), while the security against Bob is quantified by the trace distance (see Ref.~\cite{helstrom,fuchs}), and the fact that the greatest fidelity for a given trace distance is achieved when $\sigma_0$ and $\sigma_1$ are mixed \cite{NielsenChuang}.

The only aspect of the definition of state targeting
(section~\ref{motivating}) that must be changed to accommodate the
possibility of purification tests for mixed states is that at step
(iii), when Alice announces a state to Bob, she must also supply
him with a message system, so that he may measure whether the
submitted and message systems together are in a purification of
the state announced.

In this section we explain the most general strategy which Alice
can employ in order to obtain non-trivial control in state
targeting between two mixed states.

In order to increase her control, it may be to Alice's advantage
to make random choices. As is well known however\cite{Mayers},
Alice loses nothing by making these choices at a \emph{quantum}
level - that is, by entangling the system $MB$ with an ancillary
system $A$ which she retains. In general therefore, she will be
preparing the system $AMB$ in an entangled state $|\Theta
^{\scriptscriptstyle AMB}\>$. The reduced density on the system
$B$ which is submitted to Bob is then $\rho =Tr_{AM}|\Theta ^{
\scriptscriptstyle AMB}\rangle \langle \Theta ^{\scriptscriptstyle
AMB}|$. Thus, in this case Alice has submitted $\rho $ to Bob.

Prior to Alice announcing to Bob which state he should test for,
she may implement a transformation on the systems $AM$ that remain
in her possession (this includes the possibility of a measurement,
since Alice loses nothing by keeping the outcome at the quantum
level). The most general transformation is described by a
completely positive trace-preserving linear map. However, by Neumark's theorem
\cite{Peres}, an equivalent transformation can always be achieved
using a unitary map and a larger ancilla. Thus, without loss of
generality, we can assume that Alice implements a unitary
transformation $U^{\scriptscriptstyle AM}_b$, where the subscript
$b=0,1$ indicates that the operation which Alice performs will
generally depend on which of the states, $\sigma_0$ or $\sigma_1$,
is the target state.

We denote by $P(\sigma_b|\rho ,U_b^{\scriptscriptstyle AM})$ the
probability that Alice passes a test for $\sigma_b$, given that
she submitted $\rho $ to Bob and employs the interaction
$U_b^{\scriptscriptstyle AM}$ prior to sending $M$ to him. Given
our description of Alice's most general strategy, we have that
 \be
 P(\sigma_b|\rho ,U_b^{\scriptscriptstyle AM})=| \langle \Theta
^{\scriptscriptstyle AMB}|U^{\scriptscriptstyle AM}_{b} \otimes
I^{\scriptscriptstyle B} |\Psi_{b}^{\scriptscriptstyle MB} \> |^2
 \label{probunveiling}\ee
Alice's maximum control is given by
\begin{equation} \label{expression4maxcontrol}
C^{\max }=\max_{\rho
}\frac{1}{2}\sum_{b=0}^1\max_{U_b^{\scriptscriptstyle
AM}}P(\sigma_b|\rho ,U_b^{\scriptscriptstyle AM})
\end{equation}

Thus, finding the maximum control that Alice can achieve requires
two optimizations: (i) For a fixed submitted $\rho $, an
optimization over the transformation she implements on the systems
remaining in her possession after she has learnt the identity of
the target state. (ii) An optimization over the initial entangled
state $|\Theta ^{\scriptscriptstyle AMB}\>$ that she should prepare,
equivalently, over the state $\rho $ that she should submit to
Bob.

\subsection{Optimal state unveiling procedure}

In this section we will assume that the state $\rho$ which Alice
has submitted is
fixed, as is the target state, $\sigma$, that defines the purification test on $MB$ to
which her system and the message system will be subjected. We are therefore performing an
optimization of the form (i) above.

When Alice succeeds at passing a test for a state $\sigma$, we say
that she has \textit{unveiled} $\sigma$.  This terminology is
suggested by the application of state targeting to bit commitment.
 We wish to determine the maximum probability with which
Alice can pass a test for $\sigma $, i.e. unveil $\sigma$, given
that she has submitted $\rho .$

By the definition of a purification test, when Alice wishes to
unveil $\sigma$, she must send a message system $M$ to Bob, who
will then test whether the composite $MB$ is in a particular
purification $|\Psi ^{ \scriptscriptstyle MB}\>$ of $\sigma $.
As it turns out, the particular purification of
$\sigma$ that is used in the test does not affect the difficulty
of the task.

Suppose for the moment that Alice does not use an ancilla, and
instead simply prepares $MB$ in the pure state $|\Phi^{\scriptscriptstyle MB}\>$ and
implements a unitary $U^{\scriptscriptstyle M}$ on $M$ alone. In this case, the
optimization problem reduces to finding the maximum probability
with which a purification on $MB$ of $\rho$  can pass the test for
being a purification on $MB$ of $ \sigma $, and the solution is
given by \emph{Uhlmann's theorem} \cite{Jozsa}, which may be
stated as follows:
\be
\max_{U^{\scriptscriptstyle M}} | \langle \Phi
^{\scriptscriptstyle MB}|U^{\scriptscriptstyle M} \otimes
I^{\scriptscriptstyle B} |\Psi ^{\scriptscriptstyle MB} \> |^2
=F(\rho ,\sigma )^{2}, \label{Uhlmanntheorem} \\
\ee
where $F(\rho ,\sigma )\equiv \mathrm{Tr}|\sqrt{\rho }\sqrt{\sigma
}|$ is the fidelity between $\sigma$ and $\rho$.

%However, there is no reason why Alice should be restricted in this
%way, in particular she may have prepared the state $|\Phi ^{
%\scriptscriptstyle AB}\>$ over more systems than just system $B$
%and the message system $M$, and these systems (which she retains)
%may be involved in the operation she performs.

Of course, there is nothing preventing Alice from making use of an
ancilla, so one might na\"{i}vely expect her to be able to do
better than this. However, it turns out that the use of an ancilla
does \emph{not} in fact give Alice more targeting power. This fact
follows from a generalization of Uhlmann's theorem; the need for
this theorem \footnote{In Ref.~\cite{SpekkensRudolphBC}, Alice's maximum control for state targeting in the context of
a bit commitment protocol was derived without sufficient generality since the possibility of making use of an ancillary systems $A$ was not considered. The
results of that paper remain unaffected, however, since this possibility does not increase the control that Alice can achieve.}
(and an initial proof of it) was pointed out to us by
Claus D\"oscher:

\noindent \textbf{Theorem 1:}
\begin{equation}
\max_{U^{\scriptscriptstyle AM}}|\langle \Theta ^{\scriptscriptstyle AMB}|U^{\scriptscriptstyle AM}\otimes I^{\scriptscriptstyle B}|\Psi
^{\scriptscriptstyle MB}\>|^{2}=F(\rho ,\sigma )^{2},  \label{theorem1}
\end{equation}
where $U^{\scriptscriptstyle AM}$ is a unitary map on $AM$, $I^{\scriptscriptstyle B}$ is the identity map on $B$,
$\rho =\mathrm{Tr}_{\scriptscriptstyle AM}|\Theta ^{\scriptscriptstyle AMB}\rangle \langle \Theta ^{\scriptscriptstyle AMB}|$ and $
\sigma =\mathrm{Tr}_{\scriptscriptstyle M}|\Psi ^{\scriptscriptstyle MB}\rangle \langle \Psi ^{\scriptscriptstyle MB}|$.

\textbf{Proof:} To see that the left hand side (LHS) of Eq.~(\ref{theorem1})
must be greater than or equal to $F(\rho ,\sigma )^{2}$, note that

\begin{eqnarray*}
\mathrm{LHS} &=&\max_{U^{\scriptscriptstyle AM}}\sum_{k}|\langle \Theta ^{\scriptscriptstyle AMB}|U^{\scriptscriptstyle AM}\otimes
I^{\scriptscriptstyle B}|\left| \chi _{k}^{\scriptscriptstyle A}\right\rangle \otimes \left| \Psi
^{\scriptscriptstyle MB}\>\right\rangle |^{2} \\
&\ge &\max_{U^{\scriptscriptstyle AM}}|\langle \Theta ^{\scriptscriptstyle AMB}|U^{\scriptscriptstyle AM}\otimes I^{\scriptscriptstyle B}|\left| \chi
^{\scriptscriptstyle A}\right\rangle \otimes \left| \Psi ^{\scriptscriptstyle MB}\>\right\rangle |^{2} \\
&=&F(\rho ,\sigma )^{2}
\end{eqnarray*}
In the first step, we have simply introduced an arbitrary resolution of
identity for $A$ in terms of the basis $\{\left|\chi_k^{\scriptscriptstyle A}\right\rangle \}$. The second step follows from the fact
that all the terms in the sum over $k$ are positive (here, $\left| \chi
^{\scriptscriptstyle A}\right\rangle $ is an arbitrary element of the basis $\left\{ \left|
\chi _{k}^{\scriptscriptstyle A}\right\rangle \right\} $). Noting that $\left| \Theta
^{\scriptscriptstyle AMB}\right\rangle $ and  $\left| \chi ^{\scriptscriptstyle A}\right\rangle \otimes \left|
\Psi ^{\scriptscriptstyle MB}\right\rangle $ are purifications of $\rho $ and $\sigma $ on $AMB,
$ Uhlmann's theorem, Eq.~(\ref{Uhlmanntheorem}), dictates that the
optimization yields $F(\rho ,\sigma )^{2}.$

To see that the LHS of Eq.~(\ref{theorem1}) must be less than or equal to $
F(\rho ,\sigma )^{2}$, we note that if the reduced density operator of $
(U^{\scriptscriptstyle AM\dag }\otimes I^{\scriptscriptstyle B})|\Theta ^{\scriptscriptstyle AMB}\>$ over $MB$ is
denoted by $W^{\scriptscriptstyle MB}$, then $\mathrm{LHS}=\max_{U^{
\scriptscriptstyle AM}}\langle \Psi ^{\scriptscriptstyle MB}|W^{
\scriptscriptstyle MB}|\Psi ^{\scriptscriptstyle MB}\rangle =\max_{U^{
\scriptscriptstyle
AM}}F(|\Psi ^{\scriptscriptstyle MB}\rangle \langle \Psi ^{
\scriptscriptstyle MB}|,W^{\scriptscriptstyle MB})$. However for \emph{any} $
U^{\scriptscriptstyle AM}$, $\text{Tr}_{M}W^{\scriptscriptstyle MB}=\rho $,
and since the fidelity is non-decreasing under the partial trace of its
arguments \cite{NielsenChuang}, we have
\begin{eqnarray}
F(|\Psi ^{\scriptscriptstyle MB}\rangle \langle \Psi ^{\scriptscriptstyle
MB}|,W^{\scriptscriptstyle MB}) &\le &F(\mathrm{Tr}_{\scriptscriptstyle M}|\Psi ^{
\scriptscriptstyle MB}\rangle \langle \Psi ^{\scriptscriptstyle MB}|,\mathrm{
Tr}_{M}W^{\scriptscriptstyle MB})  \nonumber \\
&=&F(\sigma ,\rho ).  \nonumber
\end{eqnarray}
\leavevmode\unskip\penalty9999 \hbox{}\nobreak\hfill \quad $\blacksquare $

Theorem 1 tells us that the maximum probability of unveiling
$\sigma $ given that $\rho $ was submitted is $F(\sigma ,\rho
)^{2}$. We see from this  that it is appropriate to call $F(\sigma
,\rho )^{2}$ the `transition probability' between $\sigma $ and
$\rho $, as Uhlmann has done.

\subsection{Optimal state to submit}

The results of the previous subsection were limited to the case of a fixed submitted $\rho $. To determine Alice's maximum control we
need to optimize over $\rho $, which is an optimization of form~(ii) above.

Combining Eqs.~(\ref{expression4maxcontrol}) and
(\ref{probunveiling}) with theorem 1, we find that the maximum
control given a target state drawn uniformly from $\{\sigma
_{0},\sigma _{1}\}$ is given by
\begin{eqnarray}
C^{\max } &=&\max_{\rho }\left( \frac{1}{2}F(\sigma _{0},\rho
)^{2}+\frac{1}{ 2}F(\sigma _{1},\rho )^{2}\right) \nonumber.
\end{eqnarray}
The result of the optimization over $\rho$ is
\begin{eqnarray}
C^{\max } &=&\tfrac{1}{2}(1+F(\sigma _{0},\sigma
_{1})).\label{fred}
\end{eqnarray}
The proof of this was first presented in
Ref.~\cite{SpekkensRudolphBC} (lemma 2) and was discovered
independently in Ref.~\cite{NayakShor}.  It is reproduced in
Appendix \ref{optimalcontrol}.

The optimal strategy, i.e. the one that attains (\ref{fred}),
requires Alice to initially submit a mixed state of the form
\be\label{optrho} \rho ^{\mathrm{opt}}=\frac{\sigma _{0}+\sigma
_{1}+\sqrt{\sigma _{0}}U\sqrt{ \sigma _{1}}+\sqrt{\sigma
_{1}}U^{\dag }\sqrt{\sigma _{0}}}{2\left( 1+F\left( \sigma
_{0},\sigma _{1}\right) \right) }, \ee where \be\label{optU}
U=\left| \sqrt{\sigma _{1}}\sqrt{\sigma _{0}}\right|
^{-1}\sqrt{\sigma _{0}} \sqrt{\sigma _{1}}. \ee Again, the proof is relegated to Appendix
\ref{optimalcontrol}. It is easy to verify that these results
reduce to those presented in section \ref{ST2PSmaxcontrol} for pure states.

It is interesting to note that the maximum probability of
correctly estimating which of two mixed states applies to a system
is given by $(1+D(\sigma_0,\sigma_1))/2$, where $D$ denotes the
trace distance. We have just shown that the maximum control in
state targeting between two mixed states is given by $(1+F(\sigma
_{0},\sigma _{1}))/2$. Although the trace distance and fidelity
are known to be closely related mathematical measures of
distinguishability for mixed states, the former is generally
presumed to be much better operationally motivated because of its
connection to state estimation. Our result on maximum control can
be interpreted as providing a simple operational definition of the
fidelity.

\section{Disturbance-free state targeting for two mixed states}\label{dfst2mixed}

Consider a fourth and final weak coin flipping protocol, which is
to WCF protocol 3 as WCF protocol 2 is to WCF protocol 1.

\textbf{Weak coin flipping protocol 4}
\bi
\item[1.] Alice chooses a bit $b$
uniformly from \{0,1\}, prepares the system $MB$ in the state
$|\Psi_b\>$ and sends $B$ to Bob.
\item[2.] Bob chooses a bit $b'$ uniformly from \{0,1\} and
announces it to Alice (one can think of this as Bob's guess of
$b$).
\item[3.] Alice announces $b$ and sends the system $M$ to Bob.

\textbf{If} $b'=b$, then
\item[4.] Bob returns the system $B$ to Alice, and Alice tests
the system $MB$ for being in the state $|\Psi_b\>$.

\textbf{Else} if $b'\ne b$, then
\item[4.] Bob tests the system $MB$ for being in the state $|\Psi_b\>$.

If at step (4) the system fails Alice's(Bob's) test, then
Bob(Alice) is caught cheating. Otherwise, if $b'=b$ then Bob wins,
while if $b'\ne b$, then Alice wins.
\ei
The difference from WCF protocol 3 is that when Bob guesses $b$
correctly, he must return system $B$ to Alice.  If Bob
has gained information about the state of the system, then he has
necessarily caused a disturbance to the state, and Alice has a
probability of detecting this.  Thus, the protocol offers greater
security against Bob at the expense of security against Alice
(since she is tested less frequently).

Alice's cheating strategy in this case involves state targeting
between mixed states, as in WCF protocol 3, however unlike that
protocol, if she announces a $b$ that is equal
to $b'$, she is not tested by Bob. Thus, this is an
instance of state targeting between mixed states wherein there is
an option to decline being tested. The state targeting task is the
same as the one described in section \ref{dfstatetargeting} except
that at step (iii'), if Alice announces a state to Bob, then she
must also supply him with a message system, so that he may
implement a purification test. The possibility of declining from
being tested implies that Alice can achieve a non-trivial control
while not running any risk of failing Bob's test.  In other words,
her disturbance-free control can be greater than the trivial
control. In this section, we investigate the maximum
disturbance-free control that can be achieved in state targeting
between two mixed states.

Alice's most general strategy for this sort of state targeting
involves introducing an ancilla $A$, so that $AMB$ is in an
entangled state $|\Theta^{AMB}\>$ with reduced density operator
$\rho$ on $B$. At the time of announcement, Alice must perform a
measurement on the system $AM$ which either collapses the $B$
system to the target state $\sigma_b$ (and the $MB$ system to a
purification of $\sigma_b$) in which case she can pass the test
for $\sigma_b $ with certainty, or she collapses the $B$ system to
another state, in which case she invokes her option to not be tested by Bob.

When Alice happens to collapse the state of Bob's system to
$\sigma$, we say that she has \emph{generated} the state $\sigma$.
The distinction between state unveiling and state generation is
critical to understanding the difference between achieving state
targeting and achieving disturbance-free state targeting. A state
has been unveiled whenever a test for the state has been passed. A
state has been generated only if a test for the state would
\emph{always} be passed.

We begin by looking at Alice's ability to generate $\sigma$ given
a fixed submitted $\rho$.  We then consider variations over
$\rho$.

\subsection{Optimal state generation procedure}\label{VA}

We assume that Alice has submitted a fixed state $\rho$ and is
trying to generate the state $\sigma$. As discussed in section
\ref{dfstatetargeting}, it was proven in Ref.~\cite{HJW} that by
appropriate measurements on $AM$, Alice can update her description
of $B$ according to any extremal convex decomposition of $\rho$,
that is, any convex decomposition whose elements are all pure
states. This also holds true for non-extremal convex
decompositions (whose elements may be mixed states) as is shown in
Ref. \cite{RudolphSpekkensQIC}.  Thus, Alice's maximum probability
of generating $\sigma$ is given by the maximum probability with
which the state $\sigma$ can appear in a convex decomposition of
$\rho$. The latter is fixed by the following result, communicated
to us by Michael Nielsen:

\textbf{Theorem 2:} The maximum probability with which a  state $\sigma $
can appear in a convex decomposition of a state $\rho $ is
\be
\frac{1}{\lambda ^{\max }(\rho ^{-1}\sigma )}
\ee
if the support of $\sigma $ belongs to the support of $\rho ,$ and zero
otherwise.

\textbf{Proof:}  It is well known that if the support of $\sigma $
is not contained in the support of $\rho $ then $\sigma $ does not
appear in any convex decomposition of $\rho $ \cite{HJW}. If the
support of $\sigma $ \emph{is } so contained, then there exist
decompositions of the form \be\label{bob} \rho =p\sigma +(1-p)x
\ee where $x$ is a density operator and $0\le p\le 1.$ We seek to
determine the largest possible value of $p$; the key
constraint is that $x$ be a valid density operator. Since it
follows directly from Eq.~(\ref{bob}) that $x$ has trace 1, the
constraint becomes simply that $x$ be positive. Acting on the left
and the right of Eq.~(\ref{bob}) with $\sqrt{\rho} ^{-1}$ (where the
inverse is taken over the support of $\rho ),$ we have
\be
I=p\sqrt{\rho} ^{-1}\sigma \sqrt{\rho} ^{-1}+(1-p)\sqrt{\rho}
^{-1}x\sqrt{\rho} ^{-1}.
\ee
The constraint that $x$ be
positive implies that $(1-p)\sqrt{\rho} ^{-1}x\sqrt{\rho} ^{-1}$
is positive, which in turn implies $p\sqrt{\rho} ^{-1}\sigma
\sqrt{\rho} ^{-1}\le I$. This constrains all the eigenvalues of
$p\sqrt{\rho} ^{-1}\sigma \sqrt{\rho} ^{-1}$ to be less than
$1,$ which implies that
$p\le 1/\lambda (\sqrt{\rho} ^{-1}\sigma \sqrt{\rho} ^{-1})$ where $\lambda
(X)$ denotes an eigenvalue of $X.$ The maximum achievable value
of $p$ is simply
the smallest upper bound, which is the reciprocal of the largest eigenvalue
$%
1/\lambda ^{\max }(\sqrt{\rho} ^{-1}\sigma \sqrt{\rho} ^{-1})$.
Making use of the fact that $\lambda ^{\max }(AB)=\lambda ^{\max
}(BA),$ we obtain the desired result.\qed

\textbf{Corollary }The maximum probability with which a pure state
$\left| \psi \right\rangle \left\langle \psi \right| $ can appear
in a convex decomposition of a state $\rho $ is
\be
\frac{1}{\left\langle \psi \right| \rho ^{-1}\left| \psi \right\rangle }.
\ee
This follows from the fact that
\begin{eqnarray*}
\lambda ^{\max }(\rho ^{-1}\left| \psi \right\rangle \left\langle \psi
\right| ) &=&\lambda ^{\max }(\sqrt{\rho} ^{-1}\left| \psi \right\rangle
\left\langle \psi \right| \sqrt{\rho} ^{-1}) \\
&=&\left\langle \psi \right| \rho ^{-1}\left| \psi \right\rangle ,
\end{eqnarray*}
where the last equality is implied by the fact that $\sqrt{\rho}
^{-1}\left| \psi
\right\rangle $ is an eigenvector of $\sqrt{\rho} ^{-1}\left| \psi
\right\rangle
\left\langle \psi \right| \sqrt{\rho} ^{-1}$ with eigenvalue $\left\langle
\psi
\right| \rho ^{-1}\left| \psi \right\rangle $, and the fact that
$\sqrt{\rho} ^{-1}
\left| \psi \right\rangle \left\langle \psi \right| \sqrt{\rho} ^{-1},$
being of rank 1, has only a single non-zero eigenvalue.

Besides its obvious importance for state targeting, Theorem 2 is
also an important result for the theory of entanglement
transformation.  It specifies the largest probability of
transforming a purification of $\rho$ into a purification of
$\sigma$ given access to only the purifying share of the
bi-partite system.

\subsection{Optimal state to submit}\label{VB}

To compute the maximum disturbance-free control in state targeting
between $\sigma_0$ and $\sigma_1$, it follows from Theorem 2 that
we need to evaluate
\be
C_\text{df}^{\text{max}}=\frac{1}{2}\max_{\rho }\left(
\frac{1}{\lambda ^{\max }(\rho ^{-1}\sigma _{0})}+\frac{1}{\lambda
^{\max }(\rho ^{-1}\sigma _{1})}\right).
\ee
We have only been able to obtain bounds on this quantity, the most
simple ones being: \be\label{bounds} \frac{1}{1+\sqrt{1-F(\sigma
_{0},\sigma _{1})^{2}}}\le C_\text{df}^{\text{max}}
\le 1-\frac{1}{2}%
D(\sigma _{0},\sigma _{1}), \ee where, as before,
$F(x,y)=\mathrm{Tr}|\sqrt{x}\sqrt{y}|$ is the fidelity, and
$D(x,y)=\frac{1}{2}\mathrm{Tr}|{x}-{y}|$ is the trace distance. We
point out that $C_\text{df}^{\text{max}}$ forms an operationally
motivated measure of the distance between two mixed states.

The lower bound is derived as follows. To generate $\sigma _{0}$ or $%
\sigma _{1}$ on demand, it is sufficient to generate purifications
$\left| \Psi _{0}\right\rangle ,\left| \Psi _{1}\right\rangle $ of
$\sigma _{0}$ and $\sigma _{1}$ on demand. We know that the latter
can be done with the probability given by Eq.
(\ref{maxdfcontrol}), and the largest this probability can be made
is for maximally parallel purifications, which by Uhlmann's
theorem satisfy $\left|
\left\langle \Psi _{0}|\Psi _{1}\right\rangle \right|
^{2}=F(\sigma _{0},\sigma _{1})^{2}.$ Making this substitution
yields the advertised lower bound of Eq.~(\ref{bounds}).

Note that the lower bound is saturated for pure $\sigma _{0}$ and
$\sigma _{1}$. A number of stronger results can be obtained if
$\sigma _{0}$ and $\sigma _{1}$ are confined to a two-dimensional
Hilbert space. For instance, in the special case of $\sigma _{0}$
and $\sigma _{1}$ having equal purity, we find the lower bound of
Eq.~(\ref{bounds}) is again saturated, and for $\sigma _{0}$ and $\sigma
_{1}$ commuting, one can find a simple analytic expression for the
maximum disturbance-free control.

The upper bound is more subtle. Recall that our task is to find a
state $\rho $ and two convex decompositions of $\rho ,$ one
involving $\sigma _{0}$ and the other involving $\sigma _{1}$,
\begin{eqnarray*}
\rho  &=&p_{0}\sigma _{0}+(1-p_{0})x_{0} \\
  &=&p_{1}\sigma _{1}+(1-p_{1})x_{1},
\end{eqnarray*}
such that $\frac{1}{2}(p_{0}+p_{1})$ is maximized. We first note that
\be
p_{b}=1-\frac{D(\rho ,\sigma _{b})}{D(\sigma _{b},x_b)}.
\ee
This follows from the observation that
\begin{eqnarray*}
D(\rho ,\sigma _{b}) &=&\frac{1}{2}\mathrm{Tr}\left| (p_{b}-1)\sigma
_{b}+(1-p_{b})x_{b}\right|  \\
&=&(1-p_{b})D(\sigma _{b},x_{b})
\end{eqnarray*}
Now, making use of the fact that $D(\sigma _{b},x_{b})\le 1,$ we find
\be
\frac{1}{2}(p_{0}+p_{1})\le 1-\frac{1}{2}\left( D(\rho ,\sigma _{0})+D(\rho
,\sigma _{1})\right) .
\ee
Finally, recognizing that the trace distance is a metric and thus satisfies
$%
D(a,b)+D(b,c)\ge D(a,c),$ we obtain our upper bound. This upper
bound is sometimes saturated, for instance, when the two states
are confined to a 2-d Hilbert space with one being pure and the
other being the completely mixed state.

\section{State spying}\label{statespying}

To motivate our study of state targeting, we have introduced a
variety of weak coin flipping protocols wherein Alice's cheating
strategies are instances of state targeting. Since it has not been
our intention to provide a security analysis of any particular
protocol, we have said little about Bob's optimal cheating
strategies. Nonetheless, in the protocols where there is quantum
communication from Bob to Alice (i.e. protocols 2 and 4), Bob's cheating strategies are
instances of a task that is complementary in many respects to the
task of state targeting. We shall call this task \textit{state
spying}.

We begin by contrasting state spying to state discrimination. The
latter is what occurs in the WCF protocols where Bob is not
tested. Here, a cheating Bob is faced with the following
situation. Alice has submitted a system prepared in one of a pair
of states, and Bob seeks to guess which state this was. Thus, he
seeks to maximize his information gain, specifically, his
probability of correctly estimating the state that Alice
submitted. On the other hand, in the protocols where Bob is
required to return a quantum system to Alice whenever he guesses
correctly, he is faced with a
different task: to gain information while avoiding, as best as
possible, having Alice detect this fact. Specifically, Bob seeks
to maximize his probability of correctly estimating the state and
passing Alice's test. We call this task \textit{state spying}
because the job of a spy is not simply to gain information, but to do so without the adversary being aware of this fact.

The simplest version of the state spying task proceeds as follows:
\bi
\item[(i)] Alice submits a system in some state, called the prepared state, to Bob.
\item[(ii)] Bob implements a measurement on the system(possibly trivial) and forms
an estimate of the prepared state, called the guessed state.
\item[(iii)] Bob announces the guessed state.
\item[(iv)] Bob returns the system to Alice, and Alice performs a Pass/Fail test
for the prepared state.
\ei
The possible outcomes are:
\bi
\item[(A)] Bob announces the prepared state and passes Alice's test
\item[(B)] Bob announces the prepared state and fails Alice's test
\item[(C)] Bob announces a non-prepared state and passes Alice's test
\item[(D)] Bob announces a non-prepared state and fails Alice's test
\ei
In state spying, ``success'' is to achieve outcome (A). Pursuing the spying metaphor, we call the probability of achieving outcome~ (A) the {\em intelligence}. Another probability that is of
interest is the probability of failing Alice's test, i.e. the sum
of the probabilities of outcomes (B) or (D). We call
this the {\em disturbance.}

In the introduction, we provided a brief description of how state spying is similar to state targeting. The analogy is clearer if one compares the above definition of state spying with the definition of state targeting provided in section \ref{motivating}.  Moreover, just as it is useful for security
analyses of two-party protocols to determine the
maximal control, the maximum disturbance-free control and the
optimal control-disturbance trade-off, so too is it useful to
determine the maximum intelligence, the maximum disturbance-free
intelligence, and the optimal intelligence-disturbance trade-off.

For pure states, the maximum probability of success in
disturbance-free state spying is equal to the maximum probability
of success in error-free state discrimination, since someone who
has achieved success in an error-free discrimination procedure
knows the identity of the prepared state, and consequently can
always re-prepare the state and pass a test for it with certainty.

Mixed states can be incorporated into the notion of state spying
in exactly the same manner in which they are incorporated into
state targeting. As we have shown, a test for a purification of a
mixed state is the most difficult test to pass, and consequently
is the most interesting test to consider.  In this case, the
system that Alice submits to Bob will be entangled with one that
remains in her possession. Because of this entanglement,
successful error-free state discrimination does not imply
successful disturbance-free state spying. Although it is possible
to achieve error-free discrimination of mixed states with some
probability \cite{RudolphSpekkensTurner}, having classical
knowledge of which mixed state was prepared is insufficient to
re-prepare a purification of that state.

There are clearly many interesting questions that remain to be
answered about the task of state spying.

\section{Applications and open questions}

It is obvious that the security analyses of the WCF protocols we
have presented throughout the paper require determining the
optimal degree of success that can be achieved in a variety of
state targeting tasks. This is also true for the security analysis of
% other WCF protocols \cite{???} and of
protocols for other two-party cryptographic tasks such
as strong coin flipping \cite{wcf,Ambainis,SpekkensRudolphBC,NayakShor},
partially binding and partially concealing bit
commitment \cite{Mayers,LoChau,SpekkensRudolphBC}, and bit
escrow \cite{Aharonov}.

\textit{Quantum} two-party protocols are especially significant because in
classical information theory there are no two-party protocols
which offer information-theoretic security, that is, security which relies on the laws of physics rather than on computational intractability assumptions.
In the quantum context, there exist two-party protocols that can
guarantee to each party that if they follow the honest protocol,
their opponent cannot bias the outcome against them more than a
certain amount \cite{wcf,Aharonov,Ambainis,SpekkensRudolphBC}. Moreover, there exist quantum protocols that can guarantee
to each party that if their opponent cheats, then there will be a
non-zero probability of detecting this fact\cite{HardyKent,wcf}.
We call these two types of security {\em bias-resistance}
and {\em cheat-sensitivity} respectively. Our results have bearing on both. Specifically, the maximum control specifies the
degree of bias-resistance, while the maximum disturbance-free control,
and the optimal control-disturbance trade-off specify the degree
of cheat-sensitivity that is offered by a protocol.

The problem we have considered here can be generalized in many
ways.  First, one could consider the case where the prior
probabilities of different states being the target state are
unequal. Most of our results are easily generalized to this case.
Second, one could consider different sorts of tests, in
particular, tests that are not associated one-to-one with states,
in the sense that such a test could be passed with certainty by
more than one state.  For instance, one could consider support
tests and convex decomposition tests for mixed states (defined in
section \ref{tests}). Some results on the degree of
control that can be achieved in state targeting for convex
decomposition tests can be found in Refs.~\cite{Aharanov,Ambainis,RudolphSpekkensQIC}. Finally, one could imagine that the target state (or target test) could be chosen from a set of states (tests) with more than two elements.  Answering such questions will shed light on whether various existing protocols for two-party cryptographic tasks can be improved by modifying the sort of state targeting that is faced by a cheater.

\section{Discussion}

We have presented an introduction to a quantum information
theoretic primitive we term \emph{state targeting}.  We have argued that this is a primitive task in quantum information, in the same sense that \emph{state
estimation} is a primitive task.  Indeed, we have shown that state targeting has a natural dual in the context of state estimation, namely, the task of \emph{state spying}, which occurs when the estimator is interested in gaining information about a system while minimizing the disturbance that this entails. We have solved a variety of
optimization problems associated with state targeting. In the process, we have derived a generalization of
Uhlmann's theorem and have shown how the fidelity can arise in a
natural operational context. We end by identifying which aspects of state targeting can be deemed truly quantum.

\begin{figure}
\includegraphics[width=60mm,clip=]{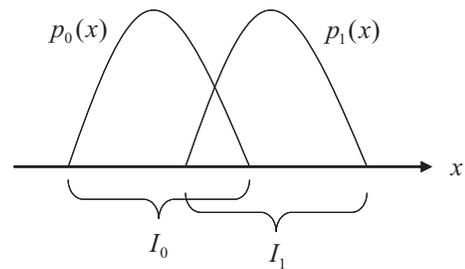}
\caption{A pair of non-orthogonal classical probability
distributions.}\label{Liouvillestates}
\end{figure}

If one adopts the view that quantum states are states of
knowledge, rather than physical states \cite{Fuchsfoundations}, what is analogous to a
quantum state in classical mechanics is a probability
distribution, and what is analogous to non-orthogonality of
quantum states is non-disjointness of probability distributions,
where two distributions are non-disjoint if there exists a
non-empty subset of the physical state space to which they both
assign non-zero probability. A pair of non-disjoint distributions
are depicted in Fig. \ref{Liouvillestates}.

By the lights of this analogy, the task of state estimation has a natural classical analogue. Specifically, the task of correctly identifying
which of a non-orthogonal pair of quantum states describes a
system is analogous to the task of correctly identifying which of
a non-disjoint pair of probability distributions describes a
classical system, that is, of identifying whether the system was prepared by drawing its physical state at random from one or another of a
pair of non-disjoint probability distributions. Just as
the quantum discrimination task cannot be accomplished with
certainty, neither can this classical discrimination task.  The
reason is that the physical state of the system has some
probability of lying in the overlap of the two probability
distributions (the region $I_0 \cap I_1$ in Fig.~\ref{Liouvillestates}), and this is consistent with both preparations.  Thus, even a measurement that completely determined the physical state of the system could not reveal which preparation was
implemented. Nonetheless, just as in quantum mechanics, there is a probability greater than 1/2 of a correct estimation, and a
non-zero probability of error-free discrimination. Specifically,
if one measures the physical state of the system, and finds it in
the region outside the overlap of the distributions (outside of $I_0 \cap I_1$), then one
knows with certainty which distribution describes the system.

Similarly, there is a classical analogue for the task of state
targeting. The target quantum state is replaced by a target
probability distribution.  Alice submits a system to Bob, then
learns the identity of the target distribution, announces a
distribution to Bob, and Bob tests the submitted system for being
described by the announced distribution.  The simplest way for Bob
to test for a distribution is to test the system for being in the
support of the distribution, that is, for being in the subset of
the physical state space that is assigned non-zero probability by
the distribution. For instance, one tests for $p_b(x)$ by
verifying whether $x$ lies in region $I_b$, depicted in Fig.
\ref{Liouvillestates}. It is now clear that Alice can achieve some
non-trivial control. For instance, if she initially submits a
system prepared according to $p_0(x)$, but the target distribution
is $p_1(x)$, then she still has some probability of passing a test
for $p_1(x)$.  Specifically, her probability is the integral of
$p_0(x)$ in the region $I_0 \cap I_1$.

For both state estimation and state targeting, there are
differences between what can be achieved in the quantum and the
classical contexts.

In the case of estimation of classical
probability distributions, one does not necessarily alter the
probability distribution by acquiring information about it, and
thus one does not alter the probability of passing subsequent
tests. For instance, if the
distribution is $p_b(x)$, then the fact that someone measures
whether $x$ is in the interval $I_0$ or not does not change the
fact that the system will be found in the interval $I_b$ when
tested. On the other hand, gaining information about which of a
pair of non-orthogonal states describes a quantum system \emph{does} influence the probability that this system will subsequently pass a test for the state in which it was initially prepared.

In the case of classical distribution targeting,
there is a strategy that allows one to pass the test for the
target state with certainty. For instance, if the target
distribution is one of $p_0(x)$ or $p_1(x)$, then Alice can simply
prepare the system in a physical state $x \in I_0 \cap I_1$ and be
certain to pass both the test for $p_0(x)$ (being in the interval
$I_0$) and the test for $p_1(x)$ (being in the interval $I_1$).
One can say that Alice achieves \textit{complete control} in this
case. On the other hand, in the quantum case, such complete
control is not possible because no quantum state can pass the test
for each of two non-orthogonal pure states with certainty.

Thus, if one takes the view that quantum states are states of
knowledge, what is surprising about quantum state estimation is
not that one cannot achieve an error-free discrimination of
non-orthogonal states with certainty but rather that one cannot
gain information without incurring a disturbance. Similarly, what
is surprising about quantum state targeting is not the possibility
of achieving a non-trivial control, but rather the impossibility
of achieving complete control, that is, the impossibility of
achieving successful state targeting with probability 1, since
this is what one would expect from examining the analogous
classical task.

Recognizing this difference between the classical and quantum
tasks is likely to be useful in devising quantum information
processing protocols that are provably superior to their classical
counterparts.

\appendix

\section{Derivation of Eqs.~(\ref{maxdfcontrol}) and (\ref{opt rho for unamb C})} \label{dfcontrol}

If the states $\rho ,$ $\left| \psi _{0}\right\rangle $ and $\left| \psi
_{1}\right\rangle $ are represented by Bloch vectors $\vec{r},$
$\hat{a}_{0},
$ and $\hat{a}_{1}$ respectively, then $1/\left\langle \psi _{b}\right| \rho
^{-1}\left| \psi _{b}\right\rangle $ $=(1-\left| \vec{r}\right| ^{2})/2(1-%
\vec{r}\cdot \hat{a}_{b}),$ and Eq.~(\ref{expression4dfcontrol})
becomes
\begin{equation}
C_{\mathrm{df}}=\frac{1}{4}(1-\left| \vec{r}\right| ^{2})\left(
\frac{1}{1-\vec{r}\cdot \hat{a}_{0}}+\frac{1}{1-\vec{r}\cdot
\hat{a}_{1}}\right) . \label{P as func of r}
\end{equation}
Extremizing $C_{\mathrm{df}}$ with respect to variations in $\vec{r},$ we
obtain
\begin{eqnarray*}
&&(1-\vec{r}\cdot \hat{a}_{0})^{2}\left( (1-\vec{r}\cdot
\hat{a}_{0})2\vec{r}%
+(1-\left| \vec{r}\right| ^{2}\hat{a}_{0}\right)  \\
&&+(1-\vec{r}\cdot \hat{a}_{1})^{2}\left( (1-\vec{r}\cdot
\hat{a}_{1})2\vec{r%
}+(1-\left| \vec{r}\right| ^{2}\hat{a}_{1}\right)  \\
&=&0
\end{eqnarray*}
One can then easily verify that the solution, $\vec{r}^{\text{opt}},$ lies
on the bisector of $\hat{a}_{0}$ and $\hat{a}_{1},$ that is,
$\vec{r}^{\text{%
opt}}\cdot (\hat{a}_{0}\times \hat{a}_{1})=0$ and $\vec{r}^{\text{opt}}\cdot
\hat{a}_{0}=\vec{r}^{\text{opt}}\cdot \hat{a}_{1},$ or equivalently,
\be
\vec{r}^{\text{opt}}=\left| \vec{r}^{\text{opt}}\right| \frac{\hat{a}_{0}+%
\hat{a}_{1}}{\left| \hat{a}_{0}+\hat{a}_{1}\right| }
\ee
and has length
\be
\left| \vec{r}^{\text{opt}}\right| =\sqrt{\frac{2}{1+\hat{a}_{0}\cdot
\hat{a}%
_{1}}}\left( 1-\sqrt{\frac{1-\hat{a}_{0}\cdot \hat{a}_{1}}{2}}\right) .
\ee
Translating these expressions from the Bloch sphere representation
to the Hilbert space representation yields Eq.~(\ref{opt rho for
unamb C}). Finally, plugging $\vec{r}^{\text{opt}}$ into Eq.~(\ref{P as func of r}) we obtain Eq.~(\ref{maxdfcontrol}).

\section{Comparison between convex decomposition tests and purification tests}
 \label{convspur}

The purification test for the mixed state $\sigma _{b}$ is a PVM measurement $
\{E_{b},I-E_{b}\}$ where
\be
E_{b}=\left| \psi _{b}\right\rangle \left\langle \psi _{b}\right|
\ee
with $\left| \psi _{b}\right\rangle $ a purification of $\sigma_b$.
Alice's
maximum control in state targeting for two mixed states, $\sigma _{0}$ and $%
\sigma _{1},$ assuming purification tests is
\begin{eqnarray*}
C_\text{pur} &=&\max_{\left| \psi \right\rangle ,U_{0},U_{1}}
\sum_{b=0}^1\tfrac{1}{2}\mathrm{Tr}\left( E_{b}(I\otimes U_{b})\left| \psi \right\rangle
\left\langle \psi \right| (I\otimes U_{b}^{\dag })\right)
  \\
&=&\max_{\left| \psi \right\rangle ,U_{0},U_{1}}
\sum_{b=0}^1\tfrac{1}{2}
\left| \< \psi _{b}\right| (I\otimes U_{b})\left| \psi \>
\right|^2 \\
&=& \max_{\rho} \left( \frac{1}{2} F(\sigma_0,\rho)^2 +
\frac{1}{2} F(\sigma_1,\rho)^2 \right).
\end{eqnarray*}
where we have used Uhlmann's theorem, Eq.~(\ref{Uhlmanntheorem}). Note that the choice of
purifications is unimportant since the probability of passing the
test does not depend on this choice; the optimization yields an
expression depending only on the reduced density operators.

Suppose $\sigma_{b}$ has a convex decomposition $\{(p_{b,k},\left|
\phi _{b,k}\right\rangle )\}$. Typically, in a convex
decomposition test Alice sends Bob a classical message containing
$b$ and $k,$ and then Bob measures the projector onto $\left| \phi
_{b,k}\right\rangle .$ However, one can think of Alice's classical
message as being conveyed by an orthogonal set of states
$\{|b,k\>\}$ of a quantum ``message'' system, and the measurement
on the system may be made conditional upon the state of this
message system. Thus, one can understand the convex decomposition
test for $\sigma_{b}$ as a PVM mesurement $\{G_{b},I-G_{b}\}$
where
\be
G_{b}=\sum_{k}\left| b,k\right\rangle \left\langle b,k\right|
\otimes \left| \phi _{b,k}\right\rangle \left\langle \phi
_{b,k}\right| .
\ee
Alice's maximum control for such a test is
\be
C_\text{con}=\max_{\left| \psi \right\rangle ,U_{0},U_{1}}
\sum_{b=0}^1\tfrac{1}{2}\mathrm{Tr}\left( G_{b}(I\otimes U_{b})\left| \psi
\right\rangle
\left\langle \psi \right| (I\otimes U_{b}^{\dag })\right)
\ee
Now, note that the state
\be
\left| \chi _{b}\right\rangle =\sum_{k}\sqrt{p_{b,k}}\left| b,k\right\rangle \left|
\phi _{b,k}\right\rangle ,
\ee
which is a purification of $\sigma_{b},$ is an eigenstate of
$G_{b}$. Thus,
\be
G_{b}=\left| \chi _{b}\right\rangle \left\langle \chi _{b}\right|
+\Gamma ,
\ee
for some positive operator $\Gamma .$

It follows that for positive $A,$ \textrm{Tr}$(G_{b}A)\ge
\mathrm{Tr}(\left| \chi _{b}\right\rangle \left\langle \chi
_{b}\right| A).$ Using this, we obtain
\be
C_\text{con}\ge \max_{\left| \psi \right\rangle ,U_{0},U_{1}}
\sum_{b=0}^1\tfrac{1}{2}\left\langle \chi _{b}\right| (I\otimes U_{b})\left| \psi
\right\rangle
\ee
But because the probability of passing a purification test is
independent of the choice of purifications, the right hand side is
simply $C_\text{pur},$ and we have
\be
C_\text{con}\ge C_\text{pur}.
\ee
Thus, state targeting with convex decomposition tests is always as
easy, or easier, than state targeting with purification tests.

\section{Derivation of Eqs.~(\ref{fred}) and (\ref{optrho})}
\label{optimalcontrol}

We begin by deriving Eq.~(\ref{fred}), i.e. that the fidelity
satisfies
\begin{equation}
\max_{\rho }\left( F^{2}\left( \rho ,\sigma \right) +F^{2}\left(
\rho ,\omega \right) \right) =1+F\left( \sigma ,\omega \right) .
\label{lemma 2 from PRA}
\end{equation}
By Uhlmann's theorem we can re-express the left-hand side of Eq.~(\ref{lemma 2 from PRA}) in terms of arbitrary purifications
$\left| \psi \right\rangle
,\left| \chi \right\rangle $ and $\left| \phi \right\rangle $ of $\rho ,$ $%
\sigma $ and $\omega ,$%
\be
\text{LHS}=\max_{\left| \psi \right\rangle }\left( \max_{U}\left|
\left\langle \chi \right| U\otimes I\left| \psi \right\rangle
\right| ^{2}+\max_{V}\left| \left\langle \phi \right| V\otimes
I\left| \psi \right\rangle \right| ^{2}\right) .
\ee
We now invert the order of the maximizations to obtain
\be
\text{LHS}=\max_{U,V}\max_{\left| \psi \right\rangle }\left(
\left| \left\langle \chi ^{\prime }|\psi \right\rangle \right|
^{2}+\left| \left\langle \phi ^{\prime }|\psi \right\rangle
\right| ^{2}\right) ,
\ee
where
\begin{eqnarray*}
\left| \chi ^{\prime }\right\rangle  &=&U^{\dag }\otimes I\left|
\chi
\right\rangle  \\
\left| \phi ^{\prime }\right\rangle  &=&V^{\dag }\otimes I\left|
\phi \right\rangle .
\end{eqnarray*}
This becomes
\begin{eqnarray}
\text{LHS} &=&\max_{U,V}\max_{\left| \psi \right\rangle }
\left| \left\langle \psi | \left( \left| \chi ^{\prime }\right\rangle
\left\langle \chi ^{\prime }\right| +\left| \phi ^{\prime
}\right\rangle \left\langle \phi
^{\prime }\right| \right) |\psi \right\rangle \right|    \label{LHS1} \\
&=&\max_{U,V}\lambda ^{\max }(\left| \chi ^{\prime }\right\rangle
\left\langle \chi ^{\prime }\right| +\left| \phi ^{\prime
}\right\rangle \left\langle \phi ^{\prime }\right| ).  \nonumber
\end{eqnarray}
where $\lambda ^{\max }(A)$ is the maximum eigenvalue of $A.$
Since
\be
\lambda ^{\max }\left( \left| \chi ^{\prime }\right\rangle
\left\langle \chi ^{\prime }\right| +\left| \phi ^{\prime
}\right\rangle \left\langle \phi ^{\prime }\right| \right)
=1+\left| \left\langle \chi ^{\prime }|\phi ^{\prime
}\right\rangle \right| .
\ee
we have
\begin{eqnarray}
\text{LHS} &=&1+\max_{U,V}\left| \left\langle \chi \right|
UV^{\dag }\otimes
I\left| \phi \right\rangle \right|   \label{LHS2} \\
&=&1+F(\omega ,\sigma ),  \nonumber
\end{eqnarray}
where in the last step we have applied Uhlmann's theorem, Eq.~(\ref{Uhlmanntheorem}).\qed

Next, we prove Eq.~(\ref{optrho}), i.e. that the density operator
$\rho $ that achieves the maximum in Eq. (\ref{lemma 2 from PRA}),
is
\be
\rho ^{\text{opt}}=\frac{\omega +\sigma +\sqrt{\sigma }U\sqrt{\omega }+\sqrt{%
\omega }U^{\dag }\sqrt{\sigma }}{2\left( 1+F\left( \sigma ,\omega
\right) \right) }, \label{rhooptappendix} \\ \ee where
\be
U=\left| \sqrt{\omega }\sqrt{\sigma }\right| ^{-1}\sqrt{\sigma
}\sqrt{\omega }.
\ee
We shall require the following lemma.

\begin{description}
\item[Lemma]  Consider two density operators, $\sigma $ and $\omega $
defined on a $d$ dimensional Hilbert space $\mathcal{H}$. Let
$\mathcal{H} ^{\prime }$ also be a $d$ dimensional Hilbert space.
The state $\left| \phi \right\rangle \in \mathcal{H}^{\prime
}\otimes \mathcal{H}$ defined by
\begin{equation}
\left| \phi \right\rangle =(I\otimes \sqrt{\omega
})\sum_{k=1}^{d}\left| e_{k}\right\rangle \left|
f_{k}\right\rangle ,  \label{phi}
\end{equation}
where $\left\{ \left| e_{k}\right\rangle \right\} _{k=1}^{d}$ is
an orthonormal basis for $\mathcal{H}^{\prime }$ and $\left\{
\left| f_{k}\right\rangle \right\} _{k=1}^{d}$ is an orthonormal
basis for $ \mathcal{H}^{\prime },$ is a purification of $\omega
.$ Furthermore, any state of the form
\begin{equation}
\left| \chi \right\rangle =\left( I\otimes \sqrt{\sigma }V\right)
\sum_{k=1}^{d}\left| e_{k}\right\rangle \left| f_{k}\right\rangle
, \label{chi}
\end{equation}
where $V$ is the unitary operator
\be
V=\left| \sqrt{\omega }\sqrt{\sigma }\right| ^{-1}\sqrt{\sigma
}\sqrt{\omega }+V_{\text{null}\left( \sqrt{\omega }\sqrt{\sigma
}\right) },
\ee
is a purification of $\sigma $ that is maximally parallel to
$\left| \phi \right\rangle .$ Note that $\left| A\right|
=\sqrt{A^{\dag }A}$, $B^{-1}$ is the inverse of $B$ on its
support, and $V_{\text{null}\left( A\right) }$ is any unitary
transformation on the null space of $A.$ Note also that an
arbitrary phase factor could be incorporated into $\left| \chi
\right\rangle .$
\end{description}

\textbf{Proof. }It is easy to verify that $\left| \phi
\right\rangle $ and $\left| \chi \right\rangle $ are indeed
purifications of $\omega $ and $ \sigma ,$ since
\begin{eqnarray*}
Tr_{\mathcal{H}}\left| \phi \right\rangle \left\langle \phi
\right| &=&\sum_{k=1}^{d}\sqrt{\omega }\left| f_{k}\right\rangle
\left\langle
f_{k}\right| \sqrt{\omega } \\
&=&\omega ,
\end{eqnarray*}
and
\begin{eqnarray*}
Tr_{\mathcal{H}}\left| \chi \right\rangle \left\langle \chi
\right| &=&\sum_{k=1}^{d}\sqrt{\sigma }V\left| f_{k}\right\rangle
\left\langle
f_{k}\right| V^{\dag }\sqrt{\sigma } \\
&=&\sqrt{\sigma }VV^{\dag }\sqrt{\sigma } \\
&=&\sigma ,
\end{eqnarray*}
where we have used the completeness of the $\{ \left|
f_{k}\right\rangle \}$ and the unitarity of $V.$ Second, $\left| \phi \right\rangle $ and $\left| \chi
\right\rangle $ are maximally parallel since
\begin{eqnarray*}
&&\left| \left\langle \phi |\chi \right\rangle \right|\\
&=&\left|
\sum_{k}\left\langle f_{k}\right| \sqrt{\omega }\sqrt{\sigma
}V\left|
f_{k}\right\rangle \right|  \\
&=&\left| Tr_{\mathcal{H}}\left( \sqrt{\sigma }V\sqrt{\omega
}\right)
\right|  \\
&=&\left| Tr_{\mathcal{H}}\left( \sqrt{\sigma }\left[ \left|
\sqrt{\omega } \sqrt{\sigma }\right| ^{-1}\sqrt{\sigma
}\sqrt{\omega }+V_{\text{null}\left( \sqrt{\omega }\sqrt{\sigma
}\right) }\right] \sqrt{\omega }\right) \right|
\\
&=&Tr_{\mathcal{H}}\left| \sqrt{\omega }\sqrt{\sigma }\right|  \\
&=&F\left( \omega ,\sigma \right) .
\end{eqnarray*}
\qed

We are now in a position to prove Eq.~(\ref{rhooptappendix}). To
begin, note that the $\left| \psi \right\rangle $ that achieves
the maximum in Eq.~(\ref{LHS1}) is simply the eigenvector of
$\left| \chi ^{\prime }\right\rangle \left\langle \chi ^{\prime
}\right| +\left| \phi ^{\prime }\right\rangle \left\langle \phi
^{\prime }\right| $ associated with the maximum eigenvalue,
\begin{equation}
\left| \psi ^{\text{opt}}\right\rangle =\frac{\left| \chi ^{\prime
}\right\rangle +\left| \phi ^{\prime }\right\rangle
}{\sqrt{2}\sqrt{1+\left| \left\langle \chi ^{\prime }|\phi
^{\prime }\right\rangle \right| }}, \label{psiopt}
\end{equation}
where we have chosen phases such that $\left\langle \chi ^{\prime
}|\phi ^{\prime }\right\rangle $ is real.

The $U$ and $V$ that achieve the maximum in Eq.~(\ref{LHS2}) are
any pair $ U^{\text{opt}},V^{\text{opt }}$such that
\be
\left| \left\langle \chi \right| U^{\text{opt}}V^{\text{opt}\dag
}\otimes I\left| \phi \right\rangle \right| =F(\omega ,\sigma ).
\ee
which implies that the states $\left| \chi ^{\prime }\right\rangle
$ and $ \left| \phi ^{\prime }\right\rangle $ that are defined in
terms of a particular $U^{\text{opt}}$ and $V^{\text{opt }}$are
maximally parallel$.$ \strut A particular pair of purifications of
$\sigma $ and $\omega $ that are maximally parallel is provided by
lemma 2. Taking Eqs.~(\ref{phi}) and (\ref{chi}) to define
$\left| \phi ^{\prime }\right\rangle $ and $\left| \chi ^{\prime
}\right\rangle ,$ we obtain
\begin{equation}
\left| \psi ^{\text{opt}}\right\rangle =\frac{(I\otimes
(\sqrt{\omega }+ \sqrt{\sigma }V))\sum_{k=1}^{d}\left|
e_{k}\right\rangle \left| f_{k}\right\rangle
}{\sqrt{2}\sqrt{1+\left| \left\langle \chi ^{\prime }|\phi
^{\prime }\right\rangle \right| }},
\end{equation}
This state allows one to achieve the maximum control. Its reduced
density operator on $\mathcal{H}$ is
\begin{eqnarray*}
\rho ^{\text{opt}} &=&\mathrm{Tr}_{\mathcal{H}^{\prime }}\left| \psi ^{\text{%
opt}}\right\rangle \left\langle \psi ^{\text{opt}}\right|  \\
&=&\frac{(\sqrt{\omega }+\sqrt{\sigma }V)(\sqrt{\omega }+V^{\dag }\sqrt{%
\sigma })}{2\left( 1+F\left( \sigma ,\omega \right) \right) } \\
&=&\frac{\omega +\sigma +\sqrt{\sigma }V\sqrt{\omega
}+\sqrt{\omega }V^{\dag }\sqrt{\sigma }}{2\left( 1+F\left( \sigma
,\omega \right) \right) }.
\end{eqnarray*}
Noting that $\sqrt{\sigma }V_{\text{null}\left( \sqrt{\omega
}\sqrt{\sigma } \right) }\sqrt{\omega }=\sqrt{\omega
}V_{\text{null}\left( \sqrt{\omega } \sqrt{\sigma }\right)
}\sqrt{\sigma }=0,$ we have the desired result.

\end{document}